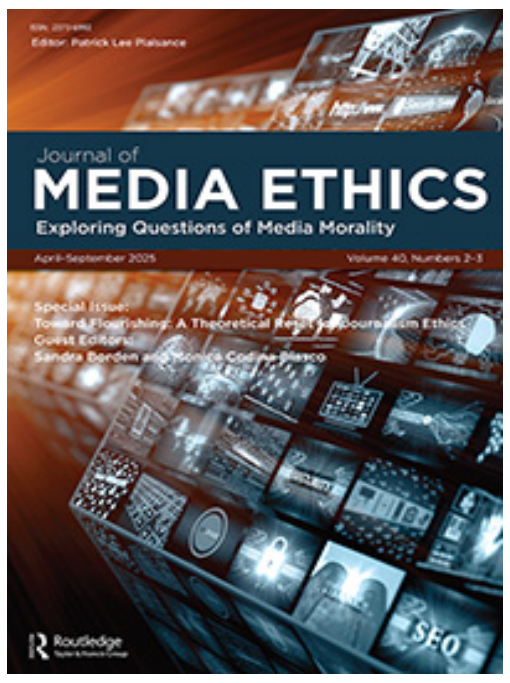

# Beyond Algorethics: Addressing the Ethical and Anthropological Challenges of AI Recommender Systems

Octavian M. Machidon

# Beyond Algorethics: Addressing the Ethical and Anthropological Challenges of AI Recommender Systems

Octavian M. Machidon

Faculty of Computer and Information Science, University of Ljubljana, Ljubljana, Slovenia

**ABSTRACT**

This paper examines the ethical and anthropological challenges posed by AI-driven recommender systems (RSs), which increasingly shape digital environments and social interactions. By curating personalized content, RSs do not merely reflect user preferences but actively construct experiences across social media, entertainment platforms, and e-commerce. Their influence raises concerns over privacy, autonomy, and mental well-being, while existing approaches such as "algorethics" – the effort to embed ethical principles into algorithmic design – remain insufficient. RSs inherently reduce human complexity to quantifiable profiles, exploit user vulnerabilities, and prioritize engagement over well-being. The paper advances a three-dimensional framework for human-centered RSs, integrating policies and regulation, interdisciplinary research, and education. These strategies are mutually reinforcing: research provides evidence for policy, policy enables safeguards and standards, and education equips users to engage critically. By connecting ethical reflection with governance and digital literacy, the paper argues that RSs can be reoriented to enhance autonomy and dignity rather than undermine them.

## Introduction

The rapid advancement of AI is reshaping the fabric of society, permeating everyday life and fundamentally altering human interactions. From social networks to digital services, AI-driven systems increasingly mediate our experiences, structuring our choices and minimizing spontaneity. Among these, AI-powered recommender systems (RSs) play a particularly pervasive role, influencing how we consume content, interact online, and make decisions. By tailoring suggestions to user preferences, RSs do not merely reflect individual choices but actively shape them, reinforcing certain behaviors while suppressing others. As a result, human engagement with the digital world becomes increasingly algorithmic, reducing direct and spontaneous experiences.

Against this backdrop, recent debates in ethics and technology studies have highlighted the broader societal risks of recommender systems. Acknowledging the longer body of work in Science and Technology Studies (STS) is essential here: Langdon Winner's classic analysis in Autonomous Technology (Winner, 1978) showed nearly five decades ago how technologies can come to define human wants and thereby reduce human control or mastery. Building on this foundation, Shoshana Zuboff's account of surveillance capitalism (Zuboff, 2019) illustrates how digital platforms compete to maximize "behavioral surplus." This imperative to extract data from human behavior entails an equally powerful imperative to connect, generating ever-denser dependencies at the human – digital interface. Paradoxically, efforts to impose control through surveillance capitalism thus create new risks

**CONTACT** Octavian M. Machidon octavian.machidon@fri.uni-lj.si Faculty of Computer and Information Science, University of Ljubljana Večna pot 113, Ljubljana 1000, Slovenia of systemic disorder, as the very structures designed for prediction and influence also heighten vulnerabilities to breakdown. In parallel, approaches such as algorethics (Benanti, 2023) and value alignment frameworks (Gabriel, 2020) stress that recommender systems must be designed not only with technical efficiency but with human dignity, privacy, and fairness at their core.

The ethical concerns raised in this body of work apply directly to AI recommender systems, which have become integral to digital life. RSs influence user engagement across various platforms, from personalized social media feeds and YouTube recommendations to TikTok's For You Page, which dominates user interaction (Narayanan, 2023). Beyond entertainment, RSs shape decisions in domains such as e-commerce (Amazon's product recommendations) and news consumption (Google's algorithmic news feeds, which impact over three billion users) (Valentine et al., 2023). Given the increasing entanglement of human activity and AI-driven recommendations – alongside proprietary and privacy concerns that limit external scrutiny – there is an urgent need for a more thorough examination of the ethical and anthropological risks these systems pose.

This paper analyzes the impact of AI recommender systems on human users, advocating for greater awareness of the ethical and anthropological challenges they present. Its central contribution is to propose a three-dimensional response – policies and regulation, interdisciplinary research, and education – that together can foster safer, more ethical, and human-centered recommender systems. By articulating this framework, the paper highlights how these three strategies are not isolated but mutually reinforcing: research provides evidence for policy, policy supports educational initiatives, and education empowers users to engage critically, creating a virtuous cycle for responsible AI development and use.

The paper proceeds as follows. Section 2 provides an overview of AI recommender systems and their key ethical challenges. Section 3 examines their anthropological impact, while Section 4 discusses the role of policy, regulation, and education in mitigating ethical risks. Section 5 critically assesses the concept of algorethics and argues for an interdisciplinary approach that goes beyond technical solutions. Next, Section 6 introduces a theoretical framework for the development of safer, human-centered AI recommender systems, and Section 7 concludes the paper.

## AI recommender systems and their ethical challenges

### *The ethical and societal impact of AI recommender systems*

AI recommender systems have become essential for managing the overwhelming volume of digital content. Traditional paradigms, such as network-based or chronological feeds, struggle to cope with the scale of information, making algorithmic recommendations indispensable (Narayanan, 2023). However, as these systems personalize user experiences, they also introduce significant ethical concerns.

At the heart of these concerns are issues of privacy, autonomy, and fairness. Recommender systems construct detailed user profiles to predict engagement, but their opacity raises transparency and accountability concerns (Milano et al., 2020). Users are often unaware of how their data is used to shape personalized content, limiting their ability to contest or control their digital experience. Moreover, AI-driven recommendations can reinforce biases, creating *filter bubbles* and *echo chambers* that restrict exposure to diverse perspectives (Alfano et al., 2021). This phenomenon not only affects individual autonomy but also has broader societal implications, particularly in shaping public discourse and reinforcing ideological polarization (Djeffal et al., 2021).

Regulatory efforts have begun to address these challenges. The European Union's *Digital Services Act* (DSA) identifies recommender systems as a potential *systemic risk*, mandating independent audits and algorithmic transparency measures (2023). Civil society organizations, such as the Irish Council for Civil Liberties, have called for stronger safeguards, citing the role of recommender systems in amplifying harmful content and exploiting user vulnerabilities. However, enforcement remains a challenge, as platforms often prioritize engagement-driven design over ethical considerations. In addition, broader critiques of surveillance capitalism (Zuboff, 2019) emphasize how the economic logic of data extraction drives many of these harms.

### Recommender systems and online radicalization

The influence of AI recommender systems extends beyond content personalization; they actively shape user behaviors, sometimes with unintended and harmful consequences. Studies have documented their role in fostering *technological seduction*, where users are incrementally guided toward extreme viewpoints without explicit intent (Alfano et al., 2018). YouTube's algorithm, for example, has been found to optimize for watch-time, inadvertently promoting conspiracy theories and extremist content by prioritizing engagement-maximizing material (Alfano et al., 2021).

Beyond individual cognitive effects, the widespread use of AI-driven recommendations raises concerns about the manipulation of public opinion. Recommender algorithms, by prioritizing content that elicits strong reactions, risk amplifying divisive and radicalizing narratives. Policymakers have recognized these threats, with proposals to enforce greater oversight on platforms whose algorithms contribute to ideological polarization and misinformation. At the same time, recent studies suggest the effects may be more nuanced, with some research finding little measurable impact of recommender systems on political beliefs (Isaac & Frenkel, 2023). Acknowledging this debate is important to avoid overstating claims.

### Mental health risks and the exploitation of user vulnerabilities

Recent investigations have highlighted how AI-driven recommendation systems exacerbate mental health issues, particularly among younger users. Amnesty International's 2023 study on TikTok revealed that the platform's recommendation algorithm disproportionately exposed users who expressed interest in mental health topics to distressing content, reinforcing harmful behavioral patterns (2023). Researchers observed a *"rabbit hole"* effect, where users – particularly teenagers – were quickly funneled into content loops romanticizing self-harm, eating disorders, and suicidal ideation.

Such patterns are not limited to TikTok. Mozilla's *RegretsReporter* project found that YouTube's recommendation algorithm was disproportionately responsible for leading users to content they later regretted engaging with, including conspiracy theories, violent imagery, and extremist narratives (McCrosky & Geurkink, 2021). These findings underscore the need for ethical intervention in AI-driven engagement strategies.

Perhaps the most tragic illustration of these risks is the case of Molly Russell, a 14-year-old who took her own life after prolonged exposure to self-harm-related content on Instagram (Djeffal et al., 2021). Her case became a catalyst for increased scrutiny of content recommendation algorithms and their impact on vulnerable users. Similar incidents have prompted advocacy groups such as *ParentsSOS* to push for stronger regulatory protections against algorithmically driven harm (2024). These examples illustrate how the design of persuasive AI intersects with public health, highlighting the importance of prevention strategies alongside regulation.

### Legal and policy responses

The growing evidence of AI recommender systems' negative impacts has prompted legal action and policy discussions worldwide. In February 2024, the city of New York filed a lawsuit against major social media platforms – including TikTok, Instagram, Facebook, Snapchat, and YouTube – alleging that their recommendation algorithms exploit young users' mental health for profit (Murphy Kelly, 2024). The lawsuit claims that these platforms contribute to increased rates of depression, anxiety, and suicidal ideation, imposing a significant financial burden on public health services. This follows broader legislative efforts in the United States to regulate AI-driven recommendation systems, with lawmakers calling for greater accountability and transparency in algorithmic decision-making.

In parallel, the *Social Media Victims Law Center* (SMVLC) has taken legal action against major platforms, seeking accountability for the harm caused by AI-driven recommendations .[1] These lawsuits reflect a growing recognition of the ethical failures of recommender systems and the urgent need for regulatory intervention.

Despite these legal and policy efforts, significant gaps remain. Current regulations primarily address transparency and risk assessment but do not mandate structural changes in AI-driven engagement strategies. Addressing these challenges requires a shift in algorithmic design priorities- one that moves beyond maximizing engagement and toward safeguarding user well-being. This shift requires close coordination between regulators, researchers, and educators, anticipating the interdisciplinary framework developed later in this paper.

The ethical challenges posed by AI recommender systems highlight the urgent need for interdisciplinary collaboration. Computer scientists, ethicists, psychologists, and policymakers must work together to develop frameworks that safeguard human dignity and autonomy. This requires regulatory measures that mandate transparency and accountability, technical research that advances bias mitigation and exposure diversity, and educational initiatives that equip users – especially young people – with the skills to navigate algorithmic environments critically. Only through such a proactive and coordinated approach can recommender systems evolve from tools that exploit vulnerabilities into technologies that support human well-being and democratic discourse.

The growing role of AI in shaping digital interactions necessitates a proactive rather than reactive approach. Without fundamental changes to how recommender systems operate, their unintended consequences will continue to impact mental health, public discourse, and individual autonomy. Ethical AI is not just a technical challenge; it is a societal imperative that demands sustained, multidisciplinary efforts to ensure that AI-driven technologies serve humanity rather than exploit its vulnerabilities.

## We become what we behold: the shaping influence of AI recommender systems

### *The transformation of human identity through AI recommendations*

Recommender systems are among the most pervasive AI technologies, mediating human interaction with digital content across social networks, online marketplaces, search engines, and even healthcare platforms. Their core function – guiding users toward content that aligns with their preferences – has led to an evolving trust dynamic between humans and AI. As these systems continuously refine their predictions, they become more than passive tools; they act as *social agents* shaping user engagement, perceptions, and even moral orientations (Omrani et al., 2022).

This influence raises profound *anthropological concerns*. By tailoring content to maximize engagement, recommender systems inherently reduce human identity to an algorithmically constructed profile based on past behavior, network dynamics, and demographic attributes (Narayanan, 2023). While this optimization enhances user experience, it simultaneously risks oversimplifying human complexity, transforming individuals into predictable data points. Within this computational framework, the human user is rendered as an input-output function, where $x$—representing behavioral data – yields a corresponding recommendation $y$.

The danger lies in the potential loss of human depth. Algorithmic efficiency prioritizes engagement, but in doing so, it compresses the richness of identity into quantifiable metrics, reinforcing behavioral patterns that may limit personal growth and intellectual diversity. The question, then, is whether these systems enhance autonomy or constrain it by shaping perception in ways that are not consciously chosen. This echoes concerns from technology studies that tools can subtly redefine human wants and capacities, diminishing mastery over one's own choices (Winner, 1978).

[1] https://socialmediavictims.org/

## Philosophical, theological, and ethical reflections on AI

The concern that technology could ultimately reshape human identity and autonomy is not new; it has been explored by philosophers and theologians for decades. Joseph Ratzinger, for instance, anticipated a future in which humanity would be compelled to define itself through computational terms, reducing individuals to numerical functions for machine processing (Benedict, 2018). He warned of a technocratic culture in which truth is equated with what is computationally possible, rather than being an objective reality beyond machine comprehension (Benedict, 2009).

Similarly, Pope Francis has become a leading voice in the global conversation on AI ethics, particularly through his advocacy of *algorethics*—the imperative to embed ethical considerations into AI development from its inception. In his 2024 World Day of Social Communications message, he asked: "How can we remain fully human and guide this cultural transformation to serve a good purpose?" (Francis, 2024). In this view, recommender systems do not merely filter information but challenge the integrity of human relationships and decision-making by reducing social experience to behavioral predictions.

## Technological determinism and AI's cognitive impact

The fear that technology shapes humanity in unintended ways is central to technological determinism, a theory suggesting that technological innovations drive cultural and societal changes beyond human control. Marshall McLuhan famously described technology as an "extension of man," asserting that new media inevitably reshape our cognition and social structures (McLuhan, 1994). His well-known adage, "The medium is the message," suggests that it is not merely the content delivered by AI that matters, but the nature of AI-driven mediation itself (McLuhan, 1962). Zuboff's notion of "surveillance capitalism" (Zuboff, 2019) further underscores how the economic logic behind these systems encourages design choices that maximize engagement and dependency rather than autonomy.

Applied to AI recommender systems, this perspective suggests that these algorithms are not simply tools for information filtering; they actively redefine human attention, engagement, and cognition. By structuring what we see and engage with, AI does not merely reflect our preferences – it reshapes them, reinforcing biases and filtering out perspectives that do not align with previous behavior.

This concern is further supported by modern scholars such as Shannon Vallor, who introduced the concept of "moral deskilling" in the digital age (Vallor, 2015). She argues that, just as automation led to the decline of manual craftsmanship, AI-driven recommendations erode our capacity for moral and intellectual discernment. When algorithms pre-select content based on engagement metrics, individuals may lose the ability to actively seek out diverse perspectives, critically evaluate information, and make independent ethical judgments.

In *The AI Mirror*, Vallor further warns that AI systems, rather than expanding human potential, often serve as mirrors reflecting and reinforcing past biases (Vallor, 2020). By continuously learning from existing human behavior, recommender systems risk amplifying societal errors rather than fostering moral and intellectual progress.

## The psychological and behavioral effects of persuasive AI

The impact of AI recommender systems on cognitive autonomy is particularly pronounced among younger users. Tristan Harris, a leading advocate for humane technology, has classified social media recommendation algorithms as "persuasive technologies" designed to manipulate user attention through behavioral nudges (2021). Features such as infinite scroll, autoplay, and algorithmic recommendations exploit cognitive biases, keeping users engaged far beyond their initial intent.

Children and adolescents, whose prefrontal cortex – the region responsible for critical thinking and impulse control – is still developing, are particularly vulnerable to these mechanisms. Studies have shown that prolonged exposure to AI-driven recommendations can lead to diminished attention spans, increased anxiety, and susceptibility to digital addiction (2021). Recent revelations from legal briefs analyzed by Haidt and Rausch (Haidt & Rausch, 2025) further expose how TikTok deliberately exploits these vulnerabilities. Internal company documents reveal that TikTok executives are acutely aware of the platform's compulsive design and its disproportionate impact on young users. Employees acknowledged that the app's recommendation algorithm is engineered to maximize engagement, even at the expense of users' mental health, with compulsive usage correlating to decreased analytical skills, impaired memory formation, and heightened anxiety. Alarmingly, TikTok's leadership has deprioritized efforts to mitigate these harms, as addressing compulsive usage was considered "not a priority for any other team."

This aligns with the concerns outlined in Section 2, where recommender systems were shown to exacerbate mental health issues by reinforcing harmful behavioral loops. For example, individuals seeking information on mental health topics may be systematically steered toward distressing content, increasing risks of self-harm and anxiety (2023). The well-documented case of Molly Russell, a teenager who tragically took her own life after prolonged exposure to self-harm-related content on Instagram, underscores the devastating consequences of engagement-driven algorithms (Djeffal et al., 2021).

Given the profound impact of AI recommender systems on human cognition and behavior, addressing these challenges demands an interdisciplinary approach. Insights from computer science, moral philosophy, psychology, theology, and law must converge to ensure that AI enhances rather than undermines human autonomy.

The key issue is not just mitigating AI's risks but shaping our relationship with it. Will we passively conform to opaque engagement algorithms, or will we design AI that fosters autonomy, critical thinking, and moral growth?

As McLuhan warns, "We become what we behold. We shape our tools, and then our tools shape us." The design of AI recommender systems is not just a technical challenge – it is an ethical and existential one that will shape human agency in the digital age.

## Policy regulations and education

Ensuring the ethical and responsible use of AI recommender systems requires a concerted effort across policy, regulation, and education. While comprehensive legislative frameworks are essential for holding platforms accountable and guiding ethical algorithmic design, education plays a critical role in equipping users – especially vulnerable demographics – with the awareness and skills to navigate digital platforms safely. Addressing these challenges demands an integrated approach in which policies mitigate systemic risks, research informs design choices, and education empowers users.

### *Regulatory efforts and policy responses*

In recent years, lawmakers have taken significant steps to regulate AI recommender systems and address their associated risks. A major milestone was the European Union's Digital Services Act (DSA), which, as of August 2023, mandates that large digital platforms assess and mitigate "systemic risks" related to their products and services. These risks include threats to fundamental rights, public discourse, mental health, and child protection. Article 34 of the DSA holds platforms accountable for preventing algorithmic amplification of harmful content and mandates greater transparency in their recommendation systems.

Beyond the EU, the U.S. Congress has also intensified its focus on AI-driven harms. Several legislative proposals in the 117th and 118th Congresses seek to curb algorithmic recommendations for children, restrict personal data usage, and enforce transparency measures on digital platforms (Busch, 2023). Some bills propose mandatory disclosure of algorithmic practices, while others advocate for non-algorithmic platform alternatives or modifications to Section 230 of the Communications Act to address liability issues stemming from algorithmic decision-making.

Yet legislation alone is insufficient without rigorous enforcement and clear evaluation standards. Ongoing investigations in the EU (Ben Mariem, 2024) and hearings in the U.S (Press & Meta, 2024). reveal that gaps persist between regulatory ambitions and industry practices, particularly regarding addictive design and child protection. This underscores the need for frameworks that move beyond transparency toward structural changes in platform design.

### *The role of education in AI literacy and digital responsibility*

While regulatory measures address corporate responsibility, education is crucial in empowering individuals to engage critically with AI recommender systems. The increasing integration of AI into digital platforms necessitates widespread AI literacy to help users recognize and navigate the ethical challenges posed by algorithmic decision-making. Educational initiatives should emphasize not only technical awareness but also the broader implications of AI on autonomy, privacy, and mental well-being.

Children and young people, as the most vulnerable demographic, require special attention. AI ethics education should be integrated into school curricula to equip students with the skills needed to engage safely and critically with AI-driven content. Teaching young users about algorithmic bias, persuasive design, and data privacy would foster digital resilience and informed decision-making.

Several organizations have taken steps to promote digital literacy and online safety. The Online Safety Hub, developed by the Safeguarding Board for Northern Ireland, provides interactive resources to educate children, parents, and educators on digital risks. In the UK, Internet Matters collaborates with policymakers and industry leaders to provide guidance on child internet safety. Their 2023 report, Online Safety in Schools, advocates for integrating digital safety education into school curricula, with a specific focus on addressing self-harm and suicide prevention (Vibert, 2023).

In the U.S., Common Sense Education – a division of the nonprofit Common Sense – offers a K–12 Digital Citizenship Curriculum designed in collaboration with Harvard's Project Zero. Covering topics such as cyberbullying, privacy, and ethical AI use, the curriculum provides educators with accessible tools to teach students about responsible digital engagement. By fostering critical thinking about AI-driven recommendations, these initiatives help young users build healthy online habits and avoid algorithmically driven manipulation.

Policy and education efforts must be informed by ongoing research into the ethical and anthropological implications of AI recommender systems. The effectiveness of regulatory frameworks depends on a nuanced understanding of how algorithms shape human behavior, influence decision-making, and contribute to societal risks. Researchers from computer science, ethics, psychology, and law must collaborate to assess the long-term consequences of AI-driven recommendations and propose safeguards that preserve human autonomy and dignity.

By combining regulatory oversight, educational initiatives, and interdisciplinary research, society can foster a balanced relationship between humans and AI-one that resists both uncritical enthusiasm and undue fear. Ensuring that recommender systems are designed and used responsibly is not merely a technical challenge but a profound moral imperative.

## Algorethics of AI recommender systems

Recognizing the ethical challenges posed by AI recommender systems necessitates integrating ethical principles into their design, an approach known as algorethics (Benanti, 2023). Algorethics seeks to ensure that AI systems align with human dignity, societal well-being, and fundamental moral values. However, defining and implementing such values in AI models remains a complex interdisciplinary challenge.

## The value alignment challenge

A fundamental initiative in this field is the Rome Call for AI Ethics, which establishes key principles – transparency, inclusion, responsibility, impartiality, reliability, security, and privacy (Pegoraro & Curzel, 2023). These principles emphasize that AI should promote human flourishing, safeguard rights, and contribute to the common good. The Rome Call represents a collaborative effort between religious institutions, policymakers, and technology leaders, fostering a human-centered AI vision.

However, translating these principles into concrete AI design methodologies presents significant challenges. AI systems must be value-aligned, meaning they should not only reflect human values but also dynamically adapt to ethical considerations. This AI alignment problem requires clear moral targets, yet current alignment strategies often lack a rigorous philosophical foundation (Hou & Green, 2023). The complexity is further exacerbated by cultural variations in ethical perspectives, making universal value alignment particularly challenging (Masso et al., 2023). As argued by recent scholarship (Machidon, 2024), consensus-based or culturally relative approaches to AI ethics risk sliding into moral relativism, leaving recommender systems without stable guidance when societal norms fluctuate. While it is unlikely that any culture would reject values such as dignity, autonomy, or fairness outright, these principles are often interpreted and prioritized differently across contexts. For example, recommender systems optimized under engagement-driven models may reflect dominant cultural or political preferences while marginalizing vulnerable groups, particularly children and adolescents. This underscores that cultural diversity is not an obstacle to be dismissed but a reminder of why alignment requires stable, universal ethical principles that can transcend shifting social consensus and provide consistent safeguards.

Gabriel (2020) highlight that alignment must consider not just AI systems and users, but also developers and society as a whole. This tetradic model underscores that recommender systems shape – and are shaped by – the broader socio-technical ecosystem. Recent work extends this perspective to advanced AI assistants, emphasizing that alignment must grapple with the divergence between revealed user preferences and long-term well-being outcomes, while also exploring how recommender systems could be designed to actively support human flourishing rather than merely mitigate harm (Gabriel et al., 2024). The challenge is therefore not only technical but also political and cultural. Several approaches – from participatory design to fairness-oriented modifications of recommendation algorithms – illustrate how alignment can be operationalized, but all remain constrained by the difficulty of balancing competing values (Fisac et al., 2020; Stray et al., 2021). Moreover, significant variations in social, cultural, and ethical values across different societies reinforce the challenge of designing universally accepted ethical AI systems (Masso et al., 2023).

Early efforts such as reinforcement learning from human feedback (RLHF) and supervised fine-tuning (SFT) provided building blocks for aligning AI with user values, but remain limited in flexibility (Lindström et al., 2024). More recent advances emphasize participatory design and human-in-the-loop approaches, which allow systems to adapt to shifting ethical expectations (Dickson, 2023; Peschl et al., 2022; Ustalov et al., 2022). These methods highlight the importance of continuous dialogue between technical design and societal oversight rather than static solutions.

## Enhancing user autonomy in recommender systems

A major ethical concern in AI-driven recommender systems is their impact on user autonomy. Personalization algorithms can create filter bubbles and echo chambers, reinforcing biases and limiting exposure to diverse perspectives (Helberger et al., 2018). AI systems, if left unchecked, may shape human preferences in ways that subtly undermine independent decision-making.

To counteract these effects, research has emphasized mechanisms such as serendipity, diversity, and randomization in recommender system design (Djeffal et al., 2021). Serendipitous recommendations expose users to novel but relevant content, helping to break self-reinforcing feedback loops. Diversity-enhancing mechanisms introduce alternative viewpoints, fostering critical thinking.

Randomization prevents deterministic filtering, ensuring users are not confined to algorithmically preselected content streams.

Djeffal et al. (Djeffal et al., 2021) argue that future recommender systems must prioritize user control to enhance self-determination. Rather than passively consuming algorithmic recommendations, users should be able to adjust system settings, choose between different recommendation models, and actively shape their content exposure. This design principle connects directly with policy and education: regulators can mandate choice architectures that promote autonomy, while digital literacy programs can teach users how to exercise such control effectively.

### *Evaluating ethical AI recommender systems*

A major gap in AI ethics research is the lack of standardized evaluation methodologies for ethical AI systems. Whether utilizing LLM-based alignment, reinforcement learning, or human-in-the-loop models, the absence of clear evaluation criteria makes it difficult to assess their real-world ethical impact.

Developing robust evaluation frameworks is essential for ensuring accountability. Such frameworks should assess the effectiveness of alignment techniques in mitigating bias and promoting fairness, the extent to which recommender systems enhance or restrict user autonomy, and the success of diversity mechanisms in broadening content exposure. They must also evaluate the ability of AI models to adapt dynamically to shifting ethical considerations, rather than embedding fixed value assumptions. In this way, evaluation methodologies not only guide AI development but also create a bridge between research and policy, helping regulators translate abstract principles into measurable standards.

In short, algorethics highlights the tension between technical innovation and ethical consistency. While value alignment methods continue to evolve, the broader challenge is to ensure that recommender systems respect autonomy, diversity, and fairness – not only through technical fixes, but also through regulatory standards and education that empower users.

## A framework supporting human-centered AI recommender systems

This paper has demonstrated that AI recommender systems wield significant influence in shaping digital environments, yet they introduce complex ethical and anthropological challenges. Their ability to personalize content has led to concerns about privacy infringement, behavioral manipulation, and exposure to harmful content – particularly among vulnerable demographics such as children and teenagers. As discussed in Section 2, numerous real-world cases illustrate the severe consequences of algorithmic recommendations, including tragic incidents of self-harm and radicalization. Furthermore, as explored in Section 3, AI-driven recommendations not only mediate digital interactions but also shape human identity, raising fundamental questions about autonomy and the reduction of human complexity to quantifiable behavioral patterns.

Addressing these challenges requires a comprehensive framework that integrates policy and regulation, interdisciplinary research, and education. The proposed framework (Figure 1) outlines a strategic approach to fostering ethical and human-centered AI recommender systems. These three dimensions are not isolated: research generates evidence for policy, policy supports educational initiatives, and education in turn creates informed users whose practices shape both design and regulation.

### *Policy and regulation: ensuring ethical AI governance*

Effective governance of AI recommender systems requires robust policies and regulatory oversight. The *Digital Services Act* (DSA) in the EU and ongoing legislative efforts in the U.S. represent important steps toward enforcing transparency and accountability in algorithmic design. However, as discussed in Section 4, current regulations remain insufficient in addressing the full scope of risks posed by AI-driven recommendations. Policymakers must move beyond general oversight and mandate the integration of ethical design principles, such as exposure diversity and user autonomy-enhancing features, directly into AI systems.

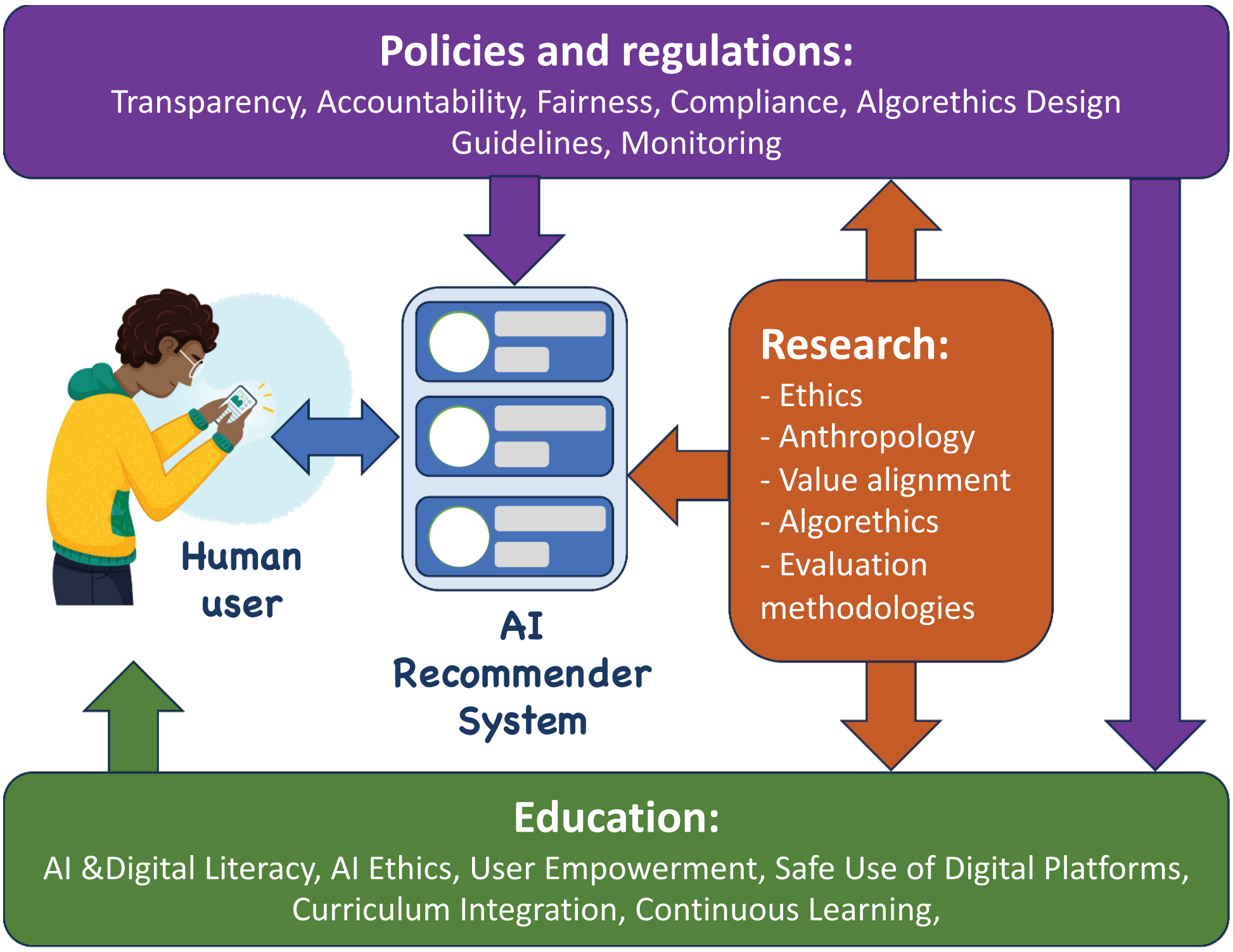

**Figure 1.** Achieving ethical, human-centered ai recommender algorithms through a concerted approach combining policies and regulations, interdisciplinary research, and education.

Additionally, regulatory frameworks should establish evaluation mechanisms to assess whether recommendation algorithms adhere to ethical standards. Currently, there is no standardized methodology for evaluating human-centered AI, leaving a critical gap in accountability. Developing clear criteria – fairness, transparency, and autonomy – would also provide researchers and educators with benchmarks for assessing platform behavior.

### *Research: advancing ethical and anthropological AI inquiry*

While technical advancements in AI recommender systems continue to enhance their predictive accuracy, research must extend beyond improving efficiency. As discussed in Section 5, ethical AI cannot be achieved solely through algorithmic refinements; it requires a fundamental shift in research priorities to investigate how these systems affect human psychology, autonomy, and societal structures.

Interdisciplinary research is particularly crucial in understanding the mechanisms of harm embedded in recommender algorithms. For example, psychologists can study compulsive usage patterns, ethicists can interrogate value alignment, and computer scientists can design protective features. These strands only gain force when connected to policy debates and educational strategies.

### *Education: empowering users in an AI-driven world*

No technical solution can fully eliminate the risks associated with AI recommender systems. While regulatory and research efforts aim to mitigate harm, education remains likely to be the most effective tool for fostering resilience against algorithmic manipulation. Given the increasing exposure of children and adolescents to AI-driven platforms, AI literacy must be integrated into school curricula from an early age.

Education should cultivate both critical awareness – recognizing algorithmic influence, bias, and persuasive design – and practical strategies for safe digital engagement. A coordinated policy approach can ensure that AI literacy becomes as routine as traditional subjects, preparing future generations to navigate algorithmic environments responsibly.

## Conclusion: toward a human-centered AI future

This paper has argued that the ethical and anthropological challenges of AI recommender systems demand more than piecemeal fixes. A human-centered future depends on aligning technical design with democratic governance, sustained research, and education that empowers citizens to navigate algorithmic environments critically.

AI should not be an autonomous force shaping human experience; it must be a tool designed and controlled by human hands. The question, then, is not whether AI recommender systems should be more efficient, but whether they should be built to serve human dignity and freedom rather than exploit vulnerabilities for engagement and profit. Meeting this challenge requires a deliberate, values-driven commitment to design systems that enhance autonomy, protect the vulnerable, and strengthen the conditions for human flourishing.

## Author contributions

CRediT: **Octavian M. Machidon:** Conceptualization, Investigation, Methodology, Writing – original draft, Writing – review & editing.

## Disclosure statement

No potential conflict of interest was reported by the author(s).

## Funding

The work was supported by The Slovenian Research and Innovation Agency [J6-60105].

## ORCID

Octavian M. Machidon http://orcid.org/0000-0003-3133-1008

## Data availability statement

We do not analyze or generate any datasets, because our work proceeds within a theoretical and mathematical approach.